\documentclass[aps,twocolumn,showpacs]{revtex4}
\usepackage{graphicx}
\begin{document}
\title{Stationary Light Pulses in Cold Atomic Media}
\author{Wen-Te Liao, Yen-Wei Lin, Thorsten Peters, Hung-Chih Chou, Jian-Siung Wang, Pei-Chen Kuan, and Ite A. Yu}
\email{yu@phys.nthu.edu.tw}
\affiliation{Department of Physics, National Tsing Hua University, Hsinchu 30013, Taiwan, Republic of China}
\date{October 28, 2008}
\begin{abstract}
Stationary light pulses (SLPs), i.e., light pulses without motion, are formed via the retrieval of stored probe pulses with two counter-propagating coupling fields. We show that there exist non-negligible hybrid Raman excitations in media of cold atoms that prohibit the SLP formation. We experimentally demonstrate a method to suppress these Raman excitations and realize SLPs in laser-cooled atoms. Our work opens the way to SLP studies in cold as well as in stationary atoms and provides a new avenue to low-light-level nonlinear optics.
\end{abstract}
\maketitle
Based on the effect of electromagnetically induced transparency (EIT) \cite{HarrisEit}, an optically dense medium can become transparent and highly dispersive for a weak probe light pulse in the presence of a strong coupling field such that the probe pulse can greatly slow down without much absorption in the medium \cite{SlowLt1, SlowLt2}. The probe pulse can even be stored in the medium by adiabatically switching off the coupling field and subsequently retrieved from the medium by the reverse process \cite{StoreLtTh, StoreLtExp1, StoreLtExp2}. Such storage and retrieval of photonic information is coherent \cite{StoreLtPhase, OurBeatNote} and provides a way to transfer a quantum state between light and matter \cite{StoreSinglePh1, StoreSinglePh2, StoreEntangle}. During the storage, there is no presence of light because the probe pulse is converted into the spin coherence of the ground states of the medium. To actually stop a light pulse while maintaining its electromagnetic component, Ref.~\cite{SLPTh1} proposed the intriguing idea of forming a  stationary light pulse (SLP) by simultaneously switching on two counter-propagating coupling fields in the retrieval process. This has been experimentally demonstrated first for a medium consisting of a hot atomic gas \cite{SLPExp1}. As SLPs significantly increase the interaction time between media and light, they are very promising for low-light-level nonlinear optics and the manipulation of photon states.

So far several theoretical papers have studied SLPs. Reference~\cite{SLPTh2} discussed the feasibility of SLPs for nonlinear optical interactions. Reference~\cite{SLPTh3} studied the diffusion and coherent control of the spatial shape of SLPs. The generation of entangled light and wavelength conversion by utilizing SLPs was explored in Refs.~\cite{SLPTh4, SLPTh5}. Reference~\cite{SLPTh6} presented a general technique to determine the solution of a multi-component SLP system in terms of dark state polaritons. How a three-dimensional Bose-Einstein condensate of stationary dark-state polaritons can be achieved was proposed in Ref.~\cite{SLPTh7}. In Ref.~\cite{SLPTh8} an analytical solution for the SLP was derived and the SLP phenomenon was studied in the case of stationary atoms such as Bose condensates and solids. They came to the conclusion, that SLPs can be created in media of stationary atoms and have much less diffusive broadening than those in media of thermal gases. 

In this Letter, we point out that there exist hybrid Raman excitations between the forward and backward fields for a medium of cold atoms which prohibit the SLP formation.  For media of hot gases these hybrid excitations can be neglected. To our knowledge, this obstacle of achieving SLPs in media of cold atoms, especially stationary atoms, has not been reported before. We then experimentally demonstrate a solution to suppress this hybrid coupling and realize SLPs in laser-cooled atoms. The terms ``hot'' and ``cold'' are here defined by the atomic motion with respect to the spatial variation of the high frequency terms of the ground-state coherence, as will be explained later.

For a three-level system consisting of two ground states $|1\rangle$ and $|2\rangle$ and an excited state $|3\rangle$ that are coupled by two light fields in $\Lambda$-type configuration, the following equations are widely used in the studies of EIT, slow light, and storage of light
\begin{eqnarray}
\label{Old31}
    \frac{\partial \rho_{31}}{\partial t} =
        \frac{i}{2} \Omega_{p} +\frac{i}{2} \Omega_{c}\rho_{21}
        -\frac{\Gamma}{2}\rho_{31}, \\
\label{Old21}
    \frac{\partial \rho_{21}}{\partial t} =
        \frac{i}{2} \Omega^{*}_{c}\rho_{31}-\gamma\rho_{21}, \\
\label{OldMaxEq}
    \frac{1}{c}\frac{\partial \Omega_p}{\partial t}
        +\frac{\partial \Omega_p}{\partial z}
         =  i \frac{\alpha \Gamma}{2L} \rho_{31}.
\end{eqnarray}
Here $\Omega_p$ and $\Omega_c$ are the Rabi frequencies of the probe and the coupling fields which drive the $|1\rangle \leftrightarrow |3\rangle$ and $|2\rangle \leftrightarrow |3\rangle$ transitions resonantly, $\rho_{31}$ and $\rho_{21}$ are the probe-transition and ground-state coherences, $\gamma$ is the ground-state relaxation rate, $\Gamma$ is the spontaneous decay rate of the excited state, and $\alpha$ and $L$ are the optical density and the length of the medium, respectively. In the case of forward- and backward-propagating probe pulses and coupling fields as needed for the creation of SLPs, we replace $\Omega_c$, $\Omega_p$, and $\rho_{31}$ by $\Omega^+_c {\rm e}^{ik z} +\Omega^-_c {\rm e}^{-ik z}$, $\Omega^+_p {\rm e}^{ik z} +\Omega^-_p {\rm e}^{-ik z}$, and $\rho^+_{31} {\rm e}^{ik z}+\rho^-_{31} {\rm e}^{-ik z}$, respectively. Here, we assume that the probe and coupling wavelengths are nearly the same, while $k$ indicates the wave vector. Equation~(\ref{Old21}) becomes
\begin{eqnarray}
   \frac{\partial \rho_{21}}{\partial t}
         = \frac{i}{2} [(\Omega^+_c)^* \rho^+_{31} +(\Omega^-_c)^* \rho^-_{31}
        +(\Omega^+_c)^* \rho^-_{31} {\rm e}^{-2ik z} \nonumber \\
        +(\Omega^-_c)^* \rho^+_{31} {\rm e}^{2ik z}] -\gamma\rho_{21}. \nonumber
\end{eqnarray}
Hence, $\rho_{21}$ can be expressed by $\rho^0_{21} + \rho^{+-}_{21} {\rm e}^{-2ik z} +\rho^{-+}_{21}{\rm e}^{2ik z}$. By neglecting terms containing ${\rm e}^{ink z}$ with $n>2$, we finally obtain the following equations:
\begin{eqnarray}
\label{Cold31Plus}
    \frac{\partial \rho^+_{31}}{\partial t}  =
        \frac{i}{2} \Omega^+_{p} +\frac{i}{2} (\Omega^+_{c}\rho^0_{21}
        +\Omega^-_{c}\rho^{-+}_{21}) -\frac{\Gamma}{2}\rho^+_{31}, \\
\label{Cold31Minus}
    \frac{\partial \rho^-_{31}}{\partial t} =
        \frac{i}{2} \Omega^-_{p} +\frac{i}{2} (\Omega^-_{c}\rho^0_{21}
        +\Omega^+_{c}\rho^{+-}_{21}) -\frac{\Gamma}{2}\rho^-_{31}, \\
\label{Cold21PM}
    \frac{\partial \rho^{+-}_{21}}{\partial t}
        = \frac{i}{2} (\Omega^+_c)^* \rho^-_{31} -\gamma_2 \rho^{+-}_{21}, \\
\label{Cold21MP}
    \frac{\partial \rho^{-+}_{21}}{\partial t}
        = \frac{i}{2} (\Omega^-_c)^* \rho^+_{31} -\gamma_2 \rho^{-+}_{21}, \\
\label{New21Zero}
    \frac{\partial \rho^0_{21}}{\partial t}
        = \frac{i}{2} [(\Omega^+_c)^* \rho^+_{31} +(\Omega^-_c)^* \rho^-_{31}]
        -\gamma_1 \rho^0_{21}, \\
\label{MaxEqPlus}
    \frac{1}{c}\frac{\partial \Omega^+_p}{\partial t}
        +\frac{\partial \Omega^+_p}{\partial z}
         =  i \frac{\alpha \Gamma}{2L} \rho^+_{31}, \\
\label{MaxEqMinus}
    \frac{1}{c}\frac{\partial \Omega^-_p}{\partial t}
        -\frac{\partial \Omega^-_p}{\partial z}
         =  i \frac{\alpha \Gamma}{2L}\rho^-_{31}.
\end{eqnarray}
Because $\rho^0_{21}$ and $\rho^{+-}_{21}$ (as well as $\rho^{-+}_{21}$) can decay differently in general, we use $\gamma_1$ and $\gamma_2$, respectively, to represent their relaxation rates.

\begin{figure}[b]
\includegraphics[width=8.25cm]{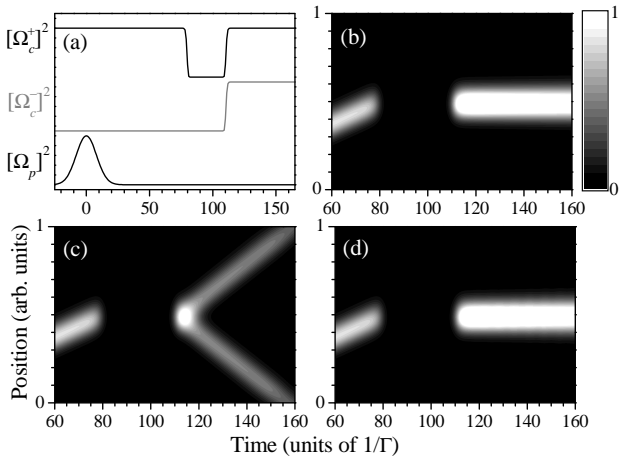}
\caption{For the timing sequence of the coupling fields and the input probe pulse depicted in (a), numerical simulations of the probe intensity ($|\Omega^+_p|^2 + |\Omega^-_p|^2$) as a function of time (horizontal axis) and position (vertical axis) are shown in (b), (c), and (d). The gray level indicates the amplitude. In (b), a medium of hot atoms is considered and Eqs.~(\ref{New21Zero})-(\ref{Hot31Minus}) are used. In (c) and (d), a medium of cold atoms is considered with $\gamma_2 = 0$ and Eqs.~(\ref{Cold31Plus})-(\ref{MaxEqMinus}) are used. The calculation parameters common to all plots are $\alpha = 2000$, $\Omega^+_c = \Omega^-_c = 3.5\Gamma$, and $\gamma_1 = 0$. The detunings in (c) and (d) are $\Delta=0$ and $1.0\Gamma$, respectively.}
\end{figure}

In room-temperature or hot media, atoms move fast and their motion destroys the high-frequency spatial variations ${\rm e}^{\pm2ik z}$. Therefore, $\gamma_2$ is very large such that $\rho^{+-}_{21}$ and $\rho^{-+}_{21}$ are negligible and
Eqs.~(\ref{Cold31Plus})-(\ref{Cold21MP}) are reduced to the already known equations for hot media
\begin{eqnarray}
\label{Hot31Plus}
    \frac{\partial \rho^+_{31}}{\partial t}  =
        \frac{i}{2} \Omega^+_{p} +\frac{i}{2} \Omega^+_{c}\rho^0_{21}
        -\frac{\Gamma}{2}\rho^+_{31}, \\
\label{Hot31Minus}
    \frac{\partial \rho^-_{31}}{\partial t} =
        \frac{i}{2} \Omega^-_{p} +\frac{i}{2} \Omega^-_{c}\rho^0_{21}
        -\frac{\Gamma}{2}\rho^-_{31}.
\end{eqnarray}
Figure~1(b) shows the numerical result for the probe pulse intensity when a SLP in a hot medium is formed for a timing of $(\Omega^+_c)^2$ and $(\Omega^-_c)^2$ depicted in Fig.~1(a). The numerical simulation is based on Eqs.~(\ref{New21Zero})-(\ref{Hot31Minus}). The probe pulse first moves into the medium in the presence of only the forward coupling field. At $t = 80\Gamma^{-1}$, the coupling field is switched off to store the probe pulse. At $t = 110\Gamma^{-1}$, both the forward and backward coupling fields are simultaneously turned on and a SLP is established.

In cold media $\rho^{+-}_{21}$ and $\rho^{-+}_{21}$ should not be neglected, i.e., $\gamma_2 \ll \Gamma$ must be allowed. Figure~1(c) shows the numerical result calculated with Eqs.~(\ref{Cold31Plus})-(\ref{MaxEqMinus}) for a medium of cold atoms with the same calculation parameters as in Fig.~1(b), except for $\gamma_2 = 0$ now. As both coupling fields of an equal intensity are turned on simultaneously at $t = 110\Gamma^{-1}$, no SLP is formed. Instead, the probe pulse splits up into two counter-propagating pulses of equal amplitude. This effect due to the hybrid Raman excitations contained in Eqs.~(\ref{Cold31Plus})-(\ref{New21Zero}) is illustrated in Fig.~2. While Figs.~2(a) and 2(b) show the coupling due to co-propagating fields, Figs.~2(c) and 2(d) show the hybrid excitations driven by counter-propagating fields. These two hybrid excitations have only to be considered for media consisting of cold atoms. Because of these excitations, it is possible, e.g., to perform slow light experiments with counter-propagating probe and coupling fields in cold media. We will show such experimental data later. As can be seen from Fig.~1(c), such hybrid Raman excitations prohibit the formation of SLPs.

\begin{figure}[b]
\includegraphics[width=8.25cm]{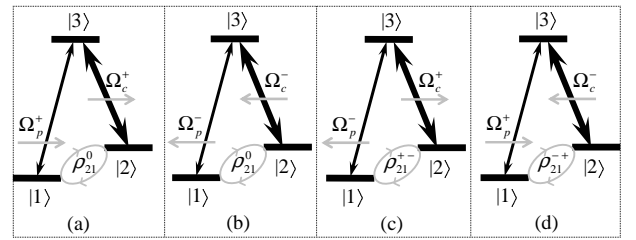}
\caption{The four Raman excitations contained in Eqs.~(\ref{Cold31Plus})-(\ref{New21Zero}) that describe the interaction between the two light fields and media of cold atoms. For media of hot atoms, only the Raman excitations in (a) and (b) need to be considered.}
\end{figure}

\begin{figure}[t]
\includegraphics[width=5.5cm]{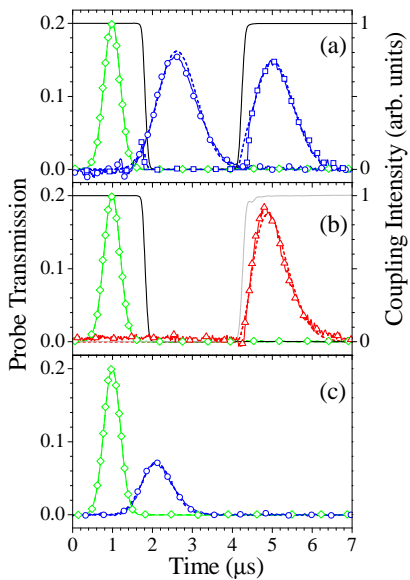}
\caption{(Color online) Experimental data (solid lines with symbols) and theoretical predictions (dashed lines without symbols) of the probe transmission through a medium of cold atoms. Equation~(\ref{Cold31Plus})-(\ref{MaxEqMinus}) are used in the calculation. (a) The forward-propagating probe pulse $|\Omega^+_p|^2$ (blue line plus circles) with the constant presence of the forward coupling field $|\Omega^+_c|^2$ (not shown) as well as the storage and retrieval of $|\Omega^+_p|^2$ (blue line plus squares) by $|\Omega^+_c|^2$ (black line). (b) Storage of $|\Omega^+_p|^2$ (not shown) by $|\Omega^+_c|^2$ (black line) and the retrieval of the backward-propagating probe pulse $|\Omega^-_p|^2$ (red line plus triangles) by the backward coupling field $|\Omega^-_c|^2$ (light gray line). (c) $|\Omega^+_p|^2$ (blue line plus circles) with the constant presence of $|\Omega^-_c|^2$ (not shown). In (a) and (b), $\alpha = 30$, $\Omega^+_c = \Omega^-_c = 0.69\Gamma$, and $\gamma_1 = 5.0\times10^{-4}\Gamma$. In (c), $\alpha = 30$, $\Omega^-_c = 0.86\Gamma$, $\gamma_2 = 0.016\Gamma$. The input probe pulse (green line plus diamonds) is drawn with its size scaled down by a factor of 0.2. Because the collection efficiencies of the two photo detectors that measured $|\Omega^+_p|^2$ and $|\Omega^-_p|^2$ were different, the red solid line representing the experimental $|\Omega^-_p|^2$ in (b) is scaled up by a factor of 1.4. The switching behaviors of the forward and backward coupling fields used in the calculation are shown in Fig.~4(a), resembling the experimental data well.}
\end{figure}

\begin{figure}[b]
\includegraphics[width=5.5cm]{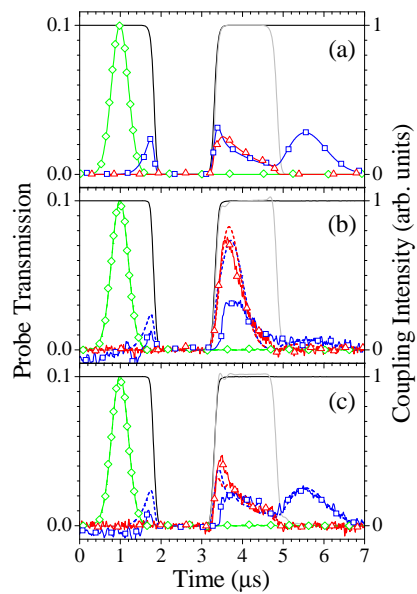}
\caption{(Color online) In (a), theoretical prediction of the probe transmission through hot atoms calculated with Eqs.~(\ref{New21Zero})-(\ref{Hot31Minus}). In (b) and (c), experimental data (solid lines with symbols) and theoretical predictions (dashed lines without symbols) of the probe transmission through a medium of cold atoms calculated with ~(\ref{Cold31Plus})-(\ref{MaxEqMinus}). After storing a probe pulse for $1.2~\mu$s both forward and backward propagating coupling fields $|\Omega^+_c|^2$ and $|\Omega^-_c|^2$ are switched on simultaneously. In the calculation, we use $\alpha = 30$, $\Omega^+_c = \Omega^-_c = 0.69\Gamma$, and $\gamma_1 = 5.0\times10^{-4}\Gamma$ as determined from the data shown in Figs.~3(a) and 3(b) and $\gamma_2 = 0.016\Gamma$ as determined from those shown in Fig.~3(c). $\Delta = 0$ in (b) and $\Delta = -0.5\Gamma$ in (c). Blue lines plus squares indicate the forward-propagating probe pulse $|\Omega^+_p|^2$, red lines plus triangles the backward-propagating probe pulse $|\Omega^-_p|^2$, black lines $|\Omega^+_c|^2$, light gray lines $|\Omega^-_c|^2$, and green lines plus diamonds the input probe pulse drawn with its size scaled down by a factor of 0.1. Because the collection efficiencies of the two photo detectors that measured $|\Omega^+_p|^2$ and $|\Omega^-_p|^2$ were different, the red solid lines representing the experimental $|\Omega^-_p|^2$ in (b) and (c) are scaled up by a factor of 1.4.}
\end{figure}

To overcome this obstacle, i.e., to prevent excitation of the two hybrid processes, we present two solutions. The first and rather difficult to achieve one is to increase $\gamma_2$ in order to make $\rho^{+-}_{21}$ and $\rho^{-+}_{21}$ negligible and suppress thereby the hybrid Raman excitations. However, $\gamma_2$ is mainly determined by the atom temperature and is not easily modified without changing other experimental parameters. A more simple and efficient way to prevent the hybrid excitations is to apply a detuning between the forward and backward coupling frequencies. Because the forward- and backward-propagating probe fields are retrieved by the forward and backward coupling fields, respectively, the Raman excitations shown in Fig.~2(a) and (b) still satisfy the two-photon resonance and provide the dominant contribution to the probe pulse propagation. On the other hand, the two hybrid processes shown in Fig.~2(c) and (d) will not fulfill the two-photon resonance and are therefore negligible. With a detuning, $\Delta$, applied between the coupling fields, we make the following revisions to Eqs.~(\ref{Cold31Minus}), (\ref{Cold21PM}), and (\ref{Cold21MP}), respectively:
\begin{eqnarray}
	\frac{\Gamma}{2} \rightarrow \frac{\Gamma}{2} - i\Delta, \nonumber \\
	\gamma_2 \rightarrow \gamma_2 - i\Delta, \nonumber \\
	\gamma_2 \rightarrow \gamma_2 + i\Delta. \nonumber
\end{eqnarray}
Calculated with $\Delta = 1.0\Gamma$ (all other parameters as in Fig.~1(c)), Fig.~1(d) shows that a SLP is formed in a medium of cold atoms. The comparison of Figs.~1(b) and 1(d) shows almost no quantitative difference, i.e., hot and cold media behave the same for SLPs when a detuning between the coupling fields is applied. Reference~\cite{SLPTh8} claims that when the two coupling intensities are imbalanced, the probe field in stationary atoms splits up into two counter-propagating pulses, i.e. no quasi-SLPs whose the moving velocity is controllably set by the coupling intensity imbalance are formed. Here, we find out that the quasi-SLPs are still produced without splitting into two parts in stationary atoms as long as a detuning is applied between the coupling fields.

In the following we discuss the experiment performed to show the validity of the previous theoretical discussion. We carried out the experiment in laser-cooled $^{87}$Rb atoms. Both coupling fields drove the transition between $|5S_{1/2},F=2\rangle$ and $|5P_{3/2},F'=2\rangle$ and both probe fields the transition between $|5S_{1/2},F=1\rangle$ and $|5P_{3/2},F'=2\rangle$. All fields had $\sigma^+$ polarization. The experimental setup has been described in detail in Ref.~\cite{OurSLPExp2}. The experimental values of the optical density $\alpha$, the coupling Rabi frequencies $\Omega^+_c$ and $\Omega^-_c$, and the ground-state relaxation rate $\gamma_1$ were determined by adjusting the parameters of the numerical simulation until its results reproduced the experimental data of the probe transmission well as shown in Figs.~3(a) and 3(b). The procedure has been described in more detail in Ref.~\cite{OurSLPExp2}. The ground-state relaxation rate $\gamma_2$ was determined similarly by performing the slow light experiment for counter-propagating probe and coupling fields as shown in Fig.~3(c).

Figure~4(a) shows the theoretical prediction of a SLP created in hot atoms calculated with the experimentally determined $\alpha$, $\Omega^+_c$, $\Omega^-_c$, and $\gamma_1$ of our system. The forward-propagating probe pulse is stored in the medium at $t = 2.0~\mu$s. Both forward and backward coupling fields are switched on simultaneously at $t = 3.2~\mu$s to convert the stored coherence into a SLP. Because the optical density of the medium is not large enough, the probe signals leaks out of the medium in the forward and backward directions while the SLP is established. The SLP is converted back to a slowly propagating pulse in the forward direction by turning off the backward coupling field at $t = 5.0~\mu$s. The pulse visible for $t > 5.0~\mu$s represents the remaining energy of the initial probe pulse after a SLP duration of $1.8~\mu$s.

Figures~4(b) and 4(c) show the experimental data and theoretical predictions for the probe transmission through a medium of cold atoms. The timing of the coupling fields is the same as in Fig.~4(a). In Fig.~4(b) the forward and backward coupling frequencies are the same ($\Delta = 0$).  When both coupling fields are present for $3.2~\mu$s $\leq t \leq$ $5.0~\mu$s two counter-propagating probe pulses are leaving the medium. There is very little observable signal for $t > 5.0~\mu$s. This behavior is in agreement with the previous theoretical discussion. The hybrid Raman excitations shown in Figs.~2(c) and 2(d) prohibit a SLP formation. Instead, the probe pulse splits up into two counter-propagating pulses that leave the medium (compare Fig.~1(c)). In Fig.~4(c), a detuning of $\Delta = -0.5\Gamma$ is applied to the backward coupling field. We clearly observe the typical signature of a SLP formed in the medium when both coupling fields are present. The pulse leaving the medium in the forward direction for $t > 5.0~\mu$s represents the remaining probe energy after a SLP duration of 1.8 $\mu$s. The experimental data in Fig.~4(c) are also consistent with the theoretical predictions in Fig.~4(a) for a medium of hot atoms. By tuning both coupling frequencies away from each other in order to inhibit the hybrid Raman excitations, media of laser-cooled atoms can be used to create SLPs as in hot media.

In conclusion, we have experimentally and theoretically studied SLPs in media of cold atoms. The experimental data are in good agreement with the theoretical predictions. Our work provides a better understanding for SLP and opens the way to SLP studies in cold as well as in stationary atoms, offering new possibilities for low-light-level nonlinear optics and manipulation of photonic information.

This work was supported by the National Science Council of Taiwan under Grants No. 95-2112-M-007-039-MY3 and No. 97-2628-M-007-018.

\newpage

\begin{thebibliography}{99}
\bibitem{HarrisEit}
S. E. Harris, Phys. Today {\bf 50}, 36 (1997).
\bibitem{SlowLt1}
L. V. Hau, S. E. Harris, Z. Dutton, and C. H. Behroozi, Nature (London) {\bf 397}, 594 (1999).
\bibitem{SlowLt2}
M. M. Kash, V. A. Sautenkov, A. S. Zibrov, L. Hollberg, G. R. Welch, M. D. Lukin, Y. Rostovtsev, E. S. Fry, and M. O. Scully, Phys. Rev. Lett. {\bf 82}, 5229 (1999).
\bibitem{StoreLtTh}
M. Fleischhauer and M. D. Lukin, Phys. Rev. Lett. {\bf 84}, 5094 (2000).
\bibitem{StoreLtExp1}
C. Liu, Z. Dutton, C. H. Behroozi, and L. V. Hau, Nature (London) {\bf 409}, 490 (2001).
\bibitem{StoreLtExp2}
D. F. Phillips, A. Fleischhauer, A. Mair, R. L. Walsworth, and M. D. Lukin, Phys. Rev. Lett. {\bf 86}, 783 (2001).
\bibitem{StoreLtPhase}
A. Mair, J. Hager, D. F. Phillips, R. L. Walsworth, and M. D. Lukin, Phys. Rev. A {\bf 65}, 031802(R) (2002).
\bibitem{OurBeatNote}
Y. F. Chen, Y. C. Liu, Z. H. Tsai, S. H. Wang, and I. A. Yu, Phys. Rev. A {\bf 72}, 033812 (2005).
\bibitem{StoreSinglePh1}
T. Chaneli\`{e}re, D. N. Matsukevich, S. D. Jenkins, S.-Y. Lan, T. A. B. Kennedy, and A. Kuzmich, Nature (London) {\bf 438}, 833 (2005).
\bibitem{StoreSinglePh2}
M. D. Eisaman, A. Andr\'{e}, F. Massou, M. Fleischhauer, A. S. Zibrov, and M. D. Lukin, Nature (London) {\bf 438}, 837 (2005).
\bibitem{StoreEntangle}
K. S. Choi, H. Deng, J. Laurat, and H. J. Kimble, Nature (London) {\bf 452}, 67 (2008).
\bibitem{SLPTh1}
A. Andr\'{e} and M. D. Lukin, Phys. Rev. Lett. {\bf 89}, 143602 (2002).
\bibitem{SLPExp1}
M. Bajcsy, A. S. Zibrov, and M. D. Lukin, Nature (London) {\bf 426}, 638 (2003).
\bibitem{SLPTh2}
A. Andr\'{e}, M. Bajcsy, A. S. Zibrov, and M. D. Lukin, Phys. Rev. Lett. {\bf 94}, 063902 (2005).
\bibitem{SLPTh3}
F. E. Zimmer, A. Andr\'{e}, M. D. Lukin, and M. Fleischhauer, Opt. Commun. {\bf 264}, 441 (2006).
\bibitem{SLPTh4}
S. A. Moiseev and B. S. Ham, Phys. Rev. A {\bf 71}, 053802 (2005).
\bibitem{SLPTh5}
S. A. Moiseev and B. S. Ham, Phys. Rev. A {\bf 73}, 033812 (2006).
\bibitem{SLPTh6}
F. E. Zimmer, J. Otterbach, R. G. Unanyan, B. W. Shore, and M. Fleischhauer, Phys. Rev. A {\bf 77}, 063823 (2008).
\bibitem{SLPTh7}
M. Fleischhauer, J. Otterbach, and R. G. Unanyan, arXiv:0807.3484v1 [quant-ph] (2008).
\bibitem{SLPTh8}
K. R. Hansen and K. M{\o}lmer, Phys. Rev. A {\bf 75}, 053802 (2007).
\bibitem{OurSLPExp2}
Y. W. Lin, H. C. Chou, T. Peters, W. T. Liao, H. W. Cho, P. C. Guan, and I. A. Yu, arXiv:0810.0903v1 [physics.optics] (2008).
\end{thebibliography}
\end{document}